\documentclass[preprint,showpacs,preprintnumbers,amsmath,amssymb]{revtex4}

\usepackage{graphicx}
\usepackage{ulem}
\usepackage{dcolumn}
\usepackage{bm}
\usepackage{textcomp}

\begin{document}

\preprint{APS/123-QED}

\title{Depth Resolution of Piezoresponse Force Microscopy}

\author{Florian Johann}
\email{johann@physik.uni-bonn.de}
\author{Tobias Jungk}
\author{\'{A}kos Hoffmann}
\author{Elisabeth Soergel}

\affiliation{Institute of Physics, University of Bonn,
Wegelerstra\ss e 8, 53115 Bonn, Germany}

\author{Yongjun J. Ying}
\author{Collin L. Sones}
\author{Robert W. Eason}
\author{Sakellaris Mailis}

\affiliation{Optoelectronics Research Centre, University of
Southampton, Highfield, Southampton, SO17 1BJ, U.\,K.}

\date{\today}

\begin{abstract}
Given that a ferroelectric domain is generally a three dimensional
entity, the determination of its area as well as its depth is
mandatory for full characterization. Piezoresponse force microscopy
(PFM) is known for its ability to map the lateral dimensions of
ferroelectric domains with high accuracy. However, no depth profile
information has been readily available so far. Here, we have used
ferroelectric domains of known depth profile to determine the
dependence of the PFM response on the depth of the domain, and thus
effectively the depth resolution of PFM detection.
\end{abstract}

\pacs{77.80.Dj, 68.37.Ps, 77.84.-s, 84.37.+q}

\maketitle


During the past decade piezoresponse force microscopy (PFM) has
become a standard tool for the investigation of ferroelectric
domains~\cite{Alexe,Jun06}. This is mainly because of its ease of
use (no specific sample preparation) combined with its capability
for imaging ferroelectric domains with high lateral resolution of $<
20$\,nm~\cite{Jun08}. Furthermore, PFM is not limited to specific
crystallographic orientations of the sample, and hence ferroelectric
domains can be visualized with PFM on all faces of the
crystal~\cite{Jun09}. Being an all-purpose analytical tool, and
therefore advantageous with respect to many other relevant
techniques used for the investigation of ferroelectric
domains~\cite{Soe05}, it is often ignored that PFM produces 2D maps
only of the domain patterns. The question that arises is: up to what
depth below the surface is PFM sensitive? While some earlier
attempts at addressing this problem were performed using thin
films~\cite{Lu02,Eng04}, to date, however, there are no reports on
measurements using single crystals. Such samples are needed
therefore as they uniquely allow for a defined domain configuration,
and thus to quantitatively determine the depth resolution of PFM.


The goal of the investigations which are presented in this paper was
to obtain a direct correlation between the depth of a surface
domain~\cite{Bus02} and the corresponding contrast obtained in PFM
measurements. The first challenge was thus to fabricate a sample
with ferroelectric surface domains of known depth. A method that can
produce such domains in lithium niobate is UV laser-induced
inhibition of poling~\cite{Son08}, a brief description of which is
given here for clarity. It was found that irradiation of the +z
polar surface of lithium niobate crystals with UV laser radiation
locally increases the coercive field. Hence, a pre-irradiated area
of the crystal surface will maintain its original polarity after a
subsequent uniform electric field poling step. The depth $d_0$ of
those poling inhibited domains is of the order of a few microns,
depending on the specific UV-writing conditions, such as the
illuminating laser light (wavelength and intensity) and scan speed
used~\cite{13}. Linear ferroelectric domain tracks several mm long
were produced by scanning the crystal sample in front of the focused
laser beam.

\begin{figure}[ttt]
\includegraphics{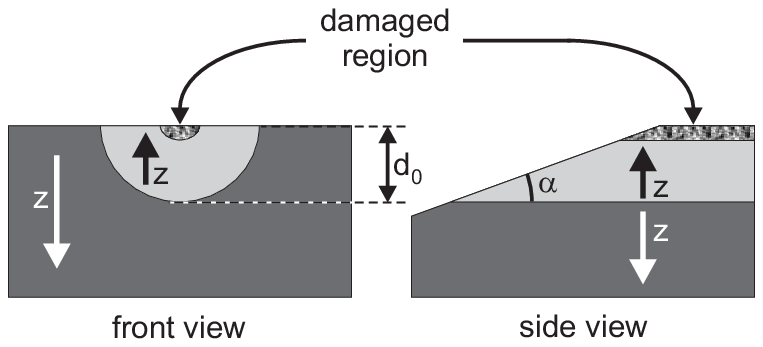}
\caption{\label{fig:johann01}
Illustration of the cross-sections of the sample used in the
experiments. A lithium niobate crystal with a stripe surface domain
(depth~$d_0$) is wedge-polished at an angle $\alpha$. At the center
of the domain the crystal is damaged due to high laser irradiation
during the fabrication process.}
\end{figure}

In order to obtain surface domains of different depth~$d_0$  the
sample was wedge-polished at a shallow angle~($\alpha = 5^{\circ}$).
For a   domain of $d_0=3$\,\textmu m depth we thereby obtained a
smooth transition from domain depths of 0 to 3\,\textmu m over a
distance of $l= d_0 / \sin \alpha \approx 35$\,\textmu m. We then
briefly etched the sample in hydrofluoric acid to enable subsequent
scanning electron microscopy (SEM) imaging.
Figure~\ref{fig:johann01} shows a schematic of the cross sections of
the wedge  polished sample. In some cases a damaged region is
observed in the centre of the poling inhibited stripe. This is a
consequence of the Gaussian profile of the irradiating UV laser beam
where the high intensity portion of the laser beam can lead to
localised melting of the surface. The melted region is then rapidly
quenched producing a polycrystalline or amorphous layer with no net
piezoelectric response.

PFM utilizes a scanning force microscope operated in contact mode
with an additional voltage applied to the tip. The imaging of
ferroelectric domains with PFM is based on the fact that
ferroelectricity implies piezoelectricity, hence mapping the
piezoelectric response of the crystal directly reflects its domain
structure. To allow sensitive readout of the piezomechanical
deformation of the material, an alternating voltage $U \sin \omega
t$ is applied to the tip and lock-in detection is used for the
measurements. A more detailed description of PFM can be found
in~\cite{Alexe,Jungk08,Jun09,Jun07}.

For the experiments we used a stand-alone scanning force microscope
(SMENA, NT-MDT, Russia). Diamond-coated tips with a nominal radius
of 50 to 70\,nm (DCP11, NT-MDT) were utilized. The voltage applied
to the tip ($\rm 5\,V_{rms}$) was directly provided by the lock-in
amplifier (SRS 830, Stanford Research Systems).

\begin{figure}[ttt]
\includegraphics{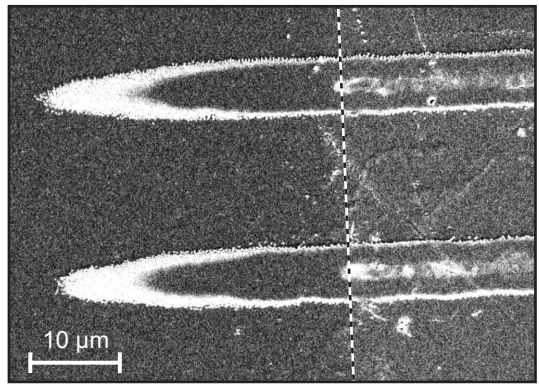}
\caption{\label{fig:johann02}
HF-etched wedge-polished sample imaged with scanning electron
microscopy. The dashed line indicates the position of the edge
caused by the wedge-polishing.}
\end{figure}
\begin{figure}[ttt]
\includegraphics{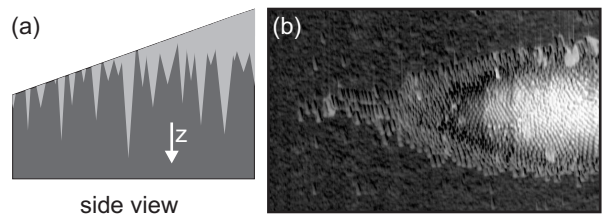}
\caption{\label{fig:johann03}
Schematic (a) of the domain configuration at the limits of the
pole-inhibited surface domain. The termination of the domain is not
sharp resulting in a grainy domain structure as it can be seen in
the PFM image (b) (image size: $15 \times 9$\,\textmu m$^2$).}
\end{figure}

Figure~\ref{fig:johann02} shows an SEM image of a wedge-polished
sample with two surface domains. In the right part of the image, the
damaged region in the centre of the domains is clearly visible.
Furthermore, a bright halo around the domains can be observed. This
feature is attributed to the imperfect boundary between the
pole-inhibited domain and the surrounding bulk domain. Because the
sample has been wedge-polished, thus gradually thinning the surface
domain, the latter appears as a composite of nano-domains at its
thinnest region, as shown in the schematic in
Fig.~\ref{fig:johann03}(a). To verify this argument we recorded
high-resolution  PFM images at the tail end of a wedge-polished
pole-inhibited domain (Fig.~\ref{fig:johann03}(b)). Obviously the
ferroelectric surface domain in this portion is no longer solid but
a composite of many nano-domains.

\begin{figure}[ttt]
\includegraphics{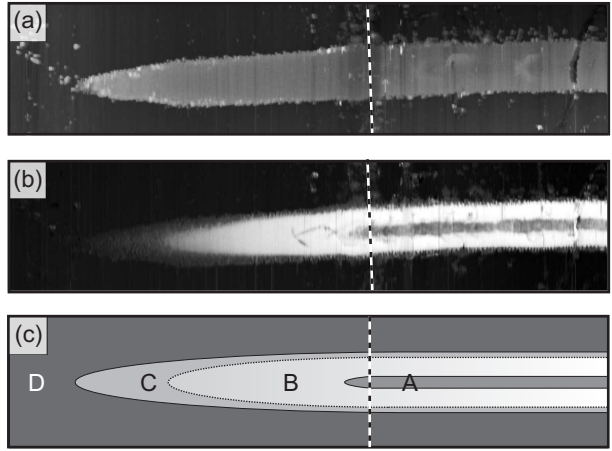}
\caption{\label{fig:johann04}
(a)~Topography and simultaneously recorded piezoresponse~(b) of the
wedge-polished sample shown in Fig.~\ref{fig:johann02}. (c) shows a
schematic of the PFM image with four distinct areas marked. A:
damaged region, B: full contrast PFM response corresponding to a
$+z$-face, C: reduced PFM response, and D: surrounding uniform
domain ($-z$-face) area. The dashed lines indicate the position of
the edge owing to wedge-polishing. For representation purposes, the
wedge has been subtracted from the topography image. Image size is
$73 \times 16$\,\textmu m$^2$.}
\end{figure}

Figure~\ref{fig:johann04} shows the results from the scanning probe
microscopy measurements of the whole wedge polished area, where
topography~(a) and piezoresponse~(b) of the sample were recorded
simultaneously. To reveal the topography of the HF-etched sample,
the slope of the wedge has been subtracted from the image by data
processing. The shape of the ferroelectric domain is the same as in
Fig.~\ref{fig:johann02}. Its maximum depth was determined to be $d_0
= l \, \sin \alpha=35.6\,\textrm{\textmu m}  \times  \sin 5^{\circ}
= 3.1$\,\textmu m. Compared with the topography, the PFM image in
(b) shows some distinct features of the surface domain. For clarity
a schematic of the PFM image is depicted in
Fig.~\ref{fig:johann04}(c). Four areas showing different amplitudes
in the PFM image are identified as follows: (A)~the stripe
associated with the central damaged region, (B)~the area with a
solid surface domain, corresponding to a $+z$-face, (C)~the part
where isolated nano-domains prevail, and (D)~the surrounding area
where the full PFM signal for the opposite orientation ($-z$-face)
is detected.

\begin{figure}[ttt]
\includegraphics{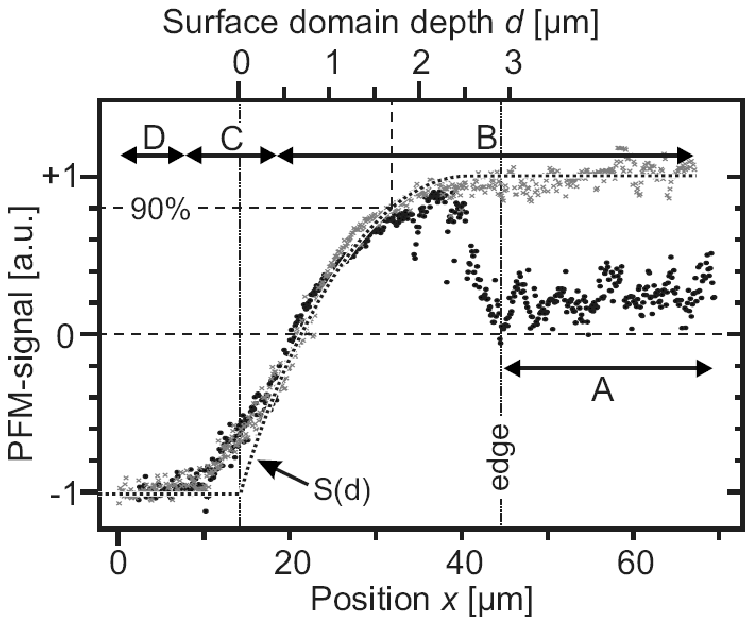}
\caption{\label{fig:johann05}
Scan-lines across the PFM image of Fig.~\ref{fig:johann04}(b), one
line passes through the damaged area (black $\bullet$) while the
other line does  not (grey $\times$). The letters indicate the areas
shown in Fig.~\ref{fig:johann04}(c).
The curve $S(d)$ is the result of the simulation.}
\end{figure}

In order to determine the depth resolution of PFM measurements the
dependence of the PFM contrast on the depth $d$ of the surface
domain must be investigated. We therefore took scan-lines along the
ferroelectric domain imaged in Fig.~\ref{fig:johann04}(b).
Figure~\ref{fig:johann05} shows two scan-lines where one passes
through the damaged area (black $\bullet$), while the other does not
(grey $\times$). The letters (A, B, C, and D) correspond to the
regions identified in Fig.~\ref{fig:johann04}(c). Note that the
presumably sharp change in the contrast between area~B and~C in
Fig.~\ref{fig:johann04}(b) cannot be observed in the slope of the
graphs in Fig.~\ref{fig:johann05}. However, these two regions can be
distinguished when comparing the noise: in region~C where we
observed the nano-domains (Fig.~\ref{fig:johann03}) the data points
fluctuate much more. In addition, at the intersection between~B
and~C the curvature of the graph changes its sign. This again is
consistent with our proposition of an uneven transition between the
surface domain and the bulk domain, leading to nano domain regions
following sample wedge-polishing. A theoretical model should
therefore only reflect part~B of the measured scan-line.


To obtain a reliable value for the depth resolution in PFM we
calculated the expected depth dependence of PFM by means of a
simplified model. We therefore approximated the spherical apex of
the tip (radius~$r$) by a point charge at the distance $r$ from the
sample surface. The resulting piezomechanical deformation was then
obtained by integrating all contributions of the sample within the
volume of the crystal experiencing the electric field from the point
charge~\cite{Jun08}. The result of our calculation can be seen in
Fig.~\ref{fig:johann05} where the curve $S(d)$ reflects favorably
the measured slope within part~B, at it is expected from the
considerations described above. The visible depth~$d_{\rm vis}$ of
PFM, i.\,e.~the depth below the crystals surface where the
contribution to the PFM signal has increased to 90\,\% of that
observed with bulk domains in a thick crystal, can be estimated to
be $d_{\rm vis}\approx 1.7$\,\textmu m. Obviously PFM cannot provide
any information about ferroelectric domains at depths $d >
1.7$\,\textmu m. In other words, surface domains with $d > d_{\rm
vis}$ can not be distinguished from bulk domains by PFM.

It would, however, be useful to establish whether this measurement
has a global rather than a material specific value. In order to
investigate to what extent this result can be generalized to other
ferroelectric materials apart from $\rm LiNbO_3$ we have considered
two extreme cases of electrostatic interaction to simulate the
interaction between the PFM tip and the surface: (a) parallel-plate
capacitor configuration and (b) the point charge model. For case
(a), the electric field inside the sample is homogeneous ($E_{\rm
z}=\rho/\varepsilon$, $\rho$ being the surface charge density), thus
while the field distribution does not depend on the material
parameters, the strength of the field, however, is a function of the
material. For the second case (b), where a point charge $q$ is
located at a distance $r$ from the sample surface, the electric
field $E_{\rm z}$ inside the crystal, normal to the sample surface
can be written as follows~\cite{Jun08}:
\begin{equation}
\label{eq:Joh2}
E_{\rm z}(x,y,z) = \frac{2q\gamma}{1+\varepsilon_{\rm eff}}
\frac{z+r}{\left[ x^2 + y^2 + \left(z+r \right)^2 \right]^{3/2}}
\end{equation}
Whereby $\gamma = \sqrt{\varepsilon_z/ \varepsilon_r}$ and
$\varepsilon_{\rm eff} = \sqrt{\varepsilon_z \varepsilon_r}$  and
$\varepsilon_z$ and $\varepsilon_r$  are the dielectric constants of
the material ($\varepsilon_z$  in $z$-direction and $\varepsilon_r$
perpendicular to $z$). As   can be seen from Eq.~\ref{eq:Joh2},
again only the amplitude of the electric field depends on the
material properties but not its spatial distribution. Although the
actual situation of the PFM tip in contact with the surface cannot
be accurately described by either case (a or b) it is expected, by
common sense interpolation of the two extreme cases calculated
above, that the actual electric field inside the sample is also
independent on the material parameters. Consequently, the visible
depth for ferroelectric domains in bulk crystals is $d_{\rm vis}
\approx 1.7\,$\textmu m, irrespective of the material. Although this
depth can be considered to be very small in terms of bulk crystals,
this value becomes important when ferroelectric domain patterns in
thin films are investigated. Not only is the thickness of the film
smaller (typically several 100\,nm) than $d_{\rm vis}$ but also the
single crystalline grains are of the order of $< 100$\,nm. PFM
images show therefore averages of several grains lying one above the
other, which is why any quantitative conclusions from PFM
measurement on such films is challenging. Note that if a lower
lateral resolution can be tolerated, the visible depth $d_{\rm vis}$
can be increased by using a tip with larger radius $r$.
Unfortunately the visible depth can not be increased by applying a
higher voltage $U \sin \omega t$ to the tip as a change of the
voltage only changes the amplitude of the signal but not the shape
of the curve.

In conclusion, we have accomplished a detailed analysis of the depth
resolution of piezoresponse force microscopy.  For the case of
lithium-niobate we have determined the visible depth to be $\approx
1.7$\,\textmu m. From basic considerations we concluded that this
depth is universal for all bulk crystals, irrespective of the
material.

\vspace{0.0cm}

\noindent \small {\bf Acknowledgments} Financial support from the
Deutsche Telekom AG, and the European Union, under the STREP
3D-Demo, is gratefully acknowledged.


\newpage

\newpage

\begin{thebibliography}{}


\bibitem{Alexe}
M.~Alexe and A.~Gruverman, eds.,
{\it Nanoscale Characterisation of Ferroelectric Materials} (Springer, Berlin; New York, 2004) 1st ed.

\bibitem{Jun06}
T.\,Jungk, A.\,Hoffmann, and E.\,Soergel, Appl.  Phys.  Lett.
\textbf{89}, 163507 (2006).

\bibitem{Jun08}
T.\,Jungk, A.\,Hoffmann, and E.\,Soergel,
New.  J.  Phys. \textbf{10}, 013019 (2008).

\bibitem{Jun09}
T.\,Jungk, A.\,Hoffmann, and E.\,Soergel,
New.  J.  Phys. \textbf{11}, 033029 (2009).

\bibitem{Soe05}
E.\,Soergel,
Appl. Phys. B: Lasers Opt. \textbf{81}, 729 (2005).

\bibitem{Lu02}
X.M.\,Lu, F.\,Schlaphof, S.\,Grafstr\"{o}m, C.\,Loppacher, and
L.M.\,Eng, Appl. Phys. Lett. \textbf{81}, 3215 (2002).

\bibitem{Eng04}
L.M. Eng, S.\,Grafstr\"{o}m, C.\,Loppacher, X.M.\,Lu, F.\,Schlaphof,
K.\,Franke, G.\,Suchaneck, and G.\,Gerlach, Integr. Ferroelectrics
\textbf{62}, 13 (2004).

\bibitem{Bus02}
A.C.\,Busacca, C.L.\,Sones, V.\,Apostolopoulos, R.W.\,Eason, and
S.\,Mailis 2002 Appl.  Phys.  Lett. \textbf{81}, 4946 (2002).

\bibitem{Son08}
C.L.\,Sones, A.C.\,Muir, Y.J.\,Ying, S.\,Mailis, R.W.\,Eason,
T.\,Jungk, A.\,Hoffmann, and E.\,Soergel, Appl. Phys.  Lett.
\textbf{92}, 072905 (2008).

\bibitem{13}
Y.J.\,Ying, C.L.\,Sones, H.\,Steigerwald, F.\,Johann, E.\,Soergel,
K.\,Buse, R.W.\,Eason, and S.\,Mailis, CLEO-Europe (2009).

\bibitem{Jungk08}
P.\,Ferraro, S.\,Grilli, and P.\,DeNatale, eds.,
{\it Ferroelectric Crystals for Photonic Applications} (Springer, Berlin; New York, 2008) 1st ed.

\bibitem{Jun07}
T.\,Jungk, A.\,Hoffmann, and E.\,Soergel,
J. Microsc. \textbf{227}, 72 (2007).

\bibitem{Ott04}
T.\,Otto, S.\,Grafstr\"{o}m, and L.\,Eng,
Ferroelectrics \textbf{303}, 149 (2004).

\bibitem{Jaz02}
M.\,Jazbinsek and M.\,Zgonik
Appl. Phys. B \textbf{74}, 407 (2002).

\bibitem{Zgo94}
M.\,Zgonik, P.\,Bernasconi, M.\,Duelli, R.\,Schlesser,
P.\,G\"{u}nter, M.H.\,Garrett, D.\,Rytz, Y.\,Zhu, and X.\,Wu, Phys.
Rev. B \textbf{50}, 5941 (1994).


\end{thebibliography}
\end{document}